\documentclass[twocolumn]{autart} 

\usepackage{amsmath} 
\usepackage{mathtools}
\usepackage{amssymb}
\usepackage{graphicx}          
\usepackage{natbib}
\usepackage{upgreek}
\usepackage{natbib}
\usepackage{upgreek}
\usepackage{xcolor}
\usepackage{booktabs} 
\usepackage[flushleft]{threeparttable}

\usepackage{comment}
\usepackage{wrapfig}
\usepackage{xcolor}

\usepackage{comment}
\usepackage{wrapfig}

\begin{document}
\begin{frontmatter}

\title{Geometric Tracking on $\mathcal{S}^{3}$ Based on Sliding Mode Control}

\thanks[footnoteinfo]{ $^{*}$Corresponding author Yu Tang, on leave from the National Autonomous University of Mexico, Mexico City, MEXICO.}

\author[UNAM]{Eduardo Espindola}\ead{eespindola@comunidad.unam.mx},
\author[UNAM]{Yu Tang$^{*}$}\ead{tang@unam.mx}               

\address[UNAM]{Ningbo Institute of Technology, Zhejiang University, Ningbo, CHINA.}  


\begin{abstract}                          
Attitude tracking on the unit sphere of dimension $3$ based on sliding mode is considered in this paper. The tangent bundle of Lagrangian dynamics that describes the rotational motion of a rigid body is first shown to be a Lie group, and then a sliding surface that emerged on it is defined.  
Next, a sliding-mode controller is designed for attitude tracking that relies on an intrinsic error defined on the Lie group. Almost global asymptotic stability of the closed loop is demonstrated using the Lyapunov analysis. Numerical simulations are included to compare the performance of the sliding mode controller designed on the Lie group with that designed in the embedding Euclidean space.
\end{abstract}
\end{frontmatter}


\section{Introduction}\label{sec1}
There has been a growing interest in designing controllers  \citep{bullo2019geometric,bullo1999tracking,spong2005controlled} and observers \citep{aghannan2003intrinsic,bonnabel2008symmetry,lageman2009gradient,zlotnik2018gradient,wang2021nonlinear} for  nonlinear systems on Lie groups due to the stronger stability and robustness that can be achieved by exploiting geometric structures of the configuration manifold such as symmetries and invariance \citep{marsden2013introduction}. 
Attitude tracking for rigid bodies using a geometric approach has been reported in \cite{markdahl2017geodesic,lee2011geometric,lee2011stable,saccon2013optimal,vang2019geometric} and references cited therein. In Refs. \cite{maithripala2006almost,maithripala2015intrinsic} an intrinsic configuration error is defined on a Lie group, and PD and PID controllers are developed with local exponential stability. These intrinsic controllers do not depend on a particular coordinate chosen, and therefore can be applied to any attitude representation in attitude tracking, e.g. rotation matrices or quaternions, with the established local exponential stability.

Sliding mode control is a widely known control methodology recognized for its ability to deal with uncertainties and external disturbances, and has been applied to design robust controller in Euclidean space \citep{utkin1978sliding,fridman2011sliding}. Its extension to nonlinear manifolds to address attitude control problems with a geometric approach has been studied in \cite{cortes2019sliding,ghasemi2020robust,meng2023second}.  Under a sliding mode controller, the error dynamics first reaches a sliding surface, designed to be invariant and to describe the desired behavior. Upon arriving the sliding surface, the error dynamics slides to the origin according to the desired behavior assigned to the sliding surface. 
Several sliding mode controllers applied to the attitude problem of a rigid body have been proposed in  \cite{crassidis1996sliding,lo1995smooth,lee2007chattering,zhu2011attitude}. One drawback of these schemes compared to the sliding mode control design with a geometric approach is that the non-Euclidean geometry of the manifold in which the system evolves is not considered in the control design, which could cause undesired behavior of the closed-loop system, such as the presence of undesired equilibria and discontinuity of the closed-loop dynamics \citep{cortes2019sliding}. 

To address these issues, in this paper we propose a sliding mode control design on the Lie group $\mathcal{S}^{3}$ using unit quaternions. The proposed sliding subgroup enables us to define a new sliding variable with geometric interpretations and allows us to design tracking controllers with stronger robustness properties. The almost global asymptotic stability of the desired equilibrium in the closed-loop system is shown by using the Lyapunov analysis. Numerical simulations are also included to compare this geometric controller with a non-geometric controller designed based on a conventional sliding variable in Euclidean space under the uncertainty in the inertia matrix and noisy measurements.  Compared to the group $SO(3)$, the unit quaternions have the advantage of easier manipulation for the implementation of control laws. The main contributions of this paper are the design of sliding mode controllers on the unit quaternion group to track a desired attitude and rigorous proof of the stability of the proposed scheme. Instead of $\mathcal S^3\times \mathbb{R}^3$  \citep{espindola2023geometric}, where $\mathbb{R}^3$ is the tangent space at the group identity, the sliding mode control design is developed directly on the state space defined by the tangent bundle $\mathcal S^3\times T_{q}\mathcal S^3$. This allows us to explore the Lagrangian dynamics for unit quaternions in the controller design. Towards this end, the state manifold of the Lagrangian dynamics based on quaternions that describe the rotational motion of a rigid body is first shown to be a Lie group under a properly defined group operation. Then a Lie subgroup embedded in the  Lie group is designed with the desired property to asymptotically stabilize the identity $\bar 1\in \mathcal{S}^{3}$ of the quaternions group. Finally, an attitude-tracking control law is devised based on the proposed sliding subgroup and the intrinsic error defined on the Lie groups.

Section \ref{sec:sec2} describes the rotational motion of a rigid body based on a 4-DOF (Degree of Freedom) Lagrangian dynamics using unit quaternions proposed in \cite{espindola2023four}. Some results and properties of the Lagrangian dynamics useful for the subsequent sliding mode control design are summarized. In Section \ref{sec:sec3}, the definition of the group operation is first defined, then the state space of the rotational Lagrangian dynamics is shown to be a Lie group. At last, a sliding surface is defined, which is shown to be embedded on the state space with desired sliding properties. In Section \ref{sec:sec4}, the evolution of the attitude error is described by a Lagrangian dynamical equation, and a tracking control law is designed. 
Section \ref{sec:sec5} provides the simulation results to illustrate the theoretical results and compare the proposed geometric controller with a non-geometric controller under parameter uncertainty and noise in the measurements. Section \ref{sec:sec6} draws the conclusions.

\subsection{Notation}

The $l_{2}$-norm is denoted by $\|x\| =\left( x^{T}x\right)^{1/2}$ for a vector $x\in\mathbb{R}^{n}$, whereas, for a matrix $A\in\mathbb{R}^{n\times m}$ its induced norm is given by $\| A\| = \left( \lambda_{\max}\left( A^{T}A\right)\right)^{1/2}$, which describes the spectral norm, where $\lambda_{\max}( A )$ and $\lambda_{\min}( A )$ respectively denote the maximum and minimum eigenvalues of $A$. The notation $\mathrm{det}(\cdot)$ and $\mathrm{tr}(\cdot)$ define the determinant and trace operation, respectively. The unit sphere and the unit ball embedded in $\mathbb{R}^{n}$ are, respectively, defined as $\mathcal{S}^{n-1}=\{ x\in\mathbb{R}^{n} \; | \; \|x\|^{2}=1 \}$ and $\mathcal{B}^{n-1}=\{ x\in\mathbb{R}^{n} \; | \; \|x\|^{2}\leq 1 \}$. The matrix $I_{n}$ describes the identity matrix of size $n\times n$. Similarly, a $n\times m$ matrix with zero entries is denoted by $0_{n\times m}$. The special orthogonal group $n$ is defined as $SO(n)=\{ R\in\mathbb{R}^{n\times n} \; |\; R^{T}R=RR^{T}=I_{n} ,\; \mathrm{det}(R)=1\}$. The Lie algebra of $SO(n)$ is $\mathfrak{so}(n) = \{ A\in\mathbb{R}^{n\times n} \;|\; A=-A^{T}\}$. Finally, the map $\cdot^{\wedge} \vcentcolon \mathbb{R}^{3} \to \mathfrak{so}(3)$ represents the cross-product operator $u^{\wedge}v = u\times v$, for all $u,v\in\mathbb{R}^{3}$.

\section{Rotational motion of the rigid body} \label{sec:sec2}

Given an inertial reference frame $\mathbf{I}$ and a body frame $\mathbf{B}$ fixed to the center of mass of the body, a rotation matrix $R\in SO(3)$ determines the attitude of the rigid body relative to the inertial frame. The rotational motion of the rigid body can be globally described by the following kinematic and dynamic equations
\begin{align}
    \dot{q} &= \frac{1}{2}J(q)\omega, \label{eq:KinQ}\\
    M\dot{\omega} &= (M\omega)^{\wedge}\omega + \tau, \label{eq:DynB}
\end{align}
where $q=\left[q_0,\vec{q}^{T}\right]^{T}\in\mathcal{S}^{3}$
is the unitary quaternion,  which is related to a rotation matrix through  Rodriguez formula $R(q) = I_{3} + 2q_0\vec{q}^{\wedge} + 2\left(\vec{q}^{\wedge}\right)^{2}$, with $q_0\in[-1,1]$ and $\vec{q}\in\mathcal{B}^{2}$ denoting the scalar and vector part of the unit quaternion, respectively.   $\omega\in\mathbb{R}^{3}$ is the angular velocity, $\tau\in\mathbb{R}^{3}$ is the torque control vector, and $M=M^{T}\in\mathbb{R}^{3\times 3}$ is a constant and positive definite inertia tensor of the rigid body, all expressed in the body frame $\mathbf{B}$. 
Map $J\vcentcolon \mathbb{R}^{4}\to \mathbb{R}^{4\times 3}$ is  defined as
\begin{align} \label{eq:J}
    J(x)&=\left[\begin{array}{c}
         -\vec{x}^{T}  \\
         x_0I_{3}+\vec{x}^{\wedge} 
    \end{array} \right], 
    \;\forall x=\left[\begin{array}{c}
         x_0 \\
         \vec{x}
    \end{array}\right] ,\; x_0\in\mathbb{R},\;\vec{x}\in\mathbb{R}^{3}.
\end{align}

Similarly, maps $L\vcentcolon \mathbb{R}^{4}\to \mathbb{R}^{4\times 3}$, and $Q,W\vcentcolon \mathbb{R}^{4}\to \mathbb{R}^{4\times 4}$, which will be used in the Lagrangian dynamics, are defined as follows
\begin{align} \label{eq:L}
    L(x)=\left[\begin{array}{c}
         -\vec{x}^{T}  \\
         x_0I_{3}-\vec{x}^{\wedge} 
    \end{array} \right], \;\forall x=\left[\begin{array}{c}
         x_0 \\
         \vec{x}
    \end{array}\right] ,\; x_0\in\mathbb{R},\;\vec{x}\in\mathbb{R}^{3} .
\end{align}
\begin{align} \label{eq:Q}
    Q(x) &= \left[ x \; J(x) \right] = \left[ \begin{array}{cc}
         x_0& -\vec{x}^{T} \\
         \vec{x}& x_0I_{3} + \vec{x}^{\wedge} 
    \end{array}\right], \\ 
    W(x)&= \left[ x \; L(x) \right] = \left[ \begin{array}{cc}
         x_0& -\vec{x}^{T} \\
         \vec{x}& x_0I_{3} - \vec{x}^{\wedge} 
    \end{array}\right]. 
\end{align}
Some useful properties of these maps are summarized in Appendix \ref{app1}.

The motion equations  \eqref{eq:KinQ}-\eqref{eq:DynB} can be described by a $4$-DOF Lagrangian dynamics proposed in \cite{espindola2023four}, given by
\begin{equation}\label{eq:EL-model}
    D(q)\ddot{q} + C(q,\dot{q})\dot{q} = \bar{\tau},
\end{equation}
where
\begin{align}
    D(q) &= J(q)MJ^T(q) + m_{0}qq^{T}, \label{eq:Dq}\\
    C(q,\dot{q}) &= -J(q)(M\omega)^{\wedge}J^T(q)-D(q)Q(\dot{q})Q^T(q), \label{eq:Cqqp}\\
    \bar{\tau} &= \frac{1}{2}J(q)\tau,\label{eq:bTau}
\end{align}
and $\lambda_{\min}(M)\leq m_{0}\leq \lambda_{\max}(M)$ a virtual inertia parameter. Note that the configuration manifold of Lagrangian dynamics \eqref{eq:EL-model} is $\mathcal{S}^3$. Moreover, the system \eqref{eq:EL-model} verifies the basic properties of Lagrangian dynamics, related to the energy conservation property, summarized in the following lemma.

\begin{lem}[\textbf{Properties of the Lagrangian system}]\label{lem2} Let $D(q)$ and $C(q,\dot{q})$ defined in \eqref{eq:Dq} and \eqref{eq:Cqqp}, respectively, then the following statements hold \citep{espindola2023four}: 
\begin{enumerate}
    \item Matrix $D(q)$ is symmetric and positive definite: 
   \begin{equation} \label{pLg1} 
   \lambda_{min}(M) I_{4}\leq D(q)\leq \lambda_{max}(M) I_{4}, \ \ \forall q\in\mathcal{S}^{3}.
    \end{equation}    

    \item Matrix $\dot{D}(q) - 2C(q,\dot{q})$ is skew-symmetric: 
\begin{equation}
x^{T} \left( \dot{D}(q) - 2C(q,\dot{q}) \right) x = 0, \ \ \forall q\in\mathcal{S}^{3}  ,\;\forall  x\in {R}^{4}. \label{pLg2}
\end{equation}    
\end{enumerate}
\end{lem}

This Lagrangian dynamics allows one to address the full attitude control problem, where the attitude may evolve on the whole group $SO(3)$, and to design control laws $\bar \tau$ based on the energy shaping methodology in Euclidean space $\mathbb{R}^{4}$ \citep{espindola2023four}. The torque applied to the rigid body is obtained by  $\tau = 2J^{T}(q)\bar{\tau}$   using Property \ref{pJ3} of matrix $J(x)$. In this paper, the geometric structure of the configuration manifold is explored, allowing one to define a sliding subgroup that inherits the geometric structure and acquires the salient features of the sliding mode through the proper design of the controller.

\section{Lie Group and Sliding Subgroup} \label{sec:sec3}

\subsection{The State Manifold as a Lie Group}
Consider the state manifold of the Lagrangian system \eqref{eq:EL-model}, given by the tangent bundle $T\mathcal{S}^{3}\vcentcolon = \mathcal{S}^{3}\times T_{q}\mathcal{S}^{3}$, $T_{q}\mathcal{S}^{3}$ being  
the tangent space at $q\in\mathcal S^3$. The following lemma establishes that $T\mathcal{S}^{3}$ is a Lie group endowed with the binary operation  
 \begin{align}\label{eq:GPct}
    g_{1}\cdot g_{2} &= \Big( Q(q_{1})q_{2} , \; Q(q_{1})p_2 + W(q_{2})p_1 \notag \\
    & \quad +\lambda Q(q_{1})\big(q_{0,2}q_{2} -\bar 1\big)+\lambda W(q_{2})\big( q_{0,1}q_{1}-\bar 1\big) \notag \\
    &\quad -\lambda \big( q_{0,12}Q(q_{1})q_{2}-\bar 1\big) \Big)\in T\mathcal{S}^{3}
\end{align}
$\forall g_{1}=(q_{1},p_1)$ and $g_{2}=(q_{2},p_2)\in T\mathcal{S}^{3}$, where $\lambda >0$ is a scalar, $\bar 1\vcentcolon=[1,0,0,0]^{T}$ is the identity element on $\mathcal{S}^{3}$, and $Q(q_{i})q_{j} = \left[ q_{0,ij},\vec{q}^{T}_{ij} \right]^{T}\in\mathcal{S}^{3}$, for $i,j=1,2,3$. Note that $Q(q_{i})q_{j}=q_{i}\otimes q_{j}\in\mathcal{S}^{3}$ corresponds to the binary operation for the Lie group $\mathcal{S}^{3}$. Moreover, with a little abuse of notation, it is denoted the triple product $q_{i}\otimes q_{j}\otimes q_{k} = \left[ q_{0,ijk},\vec{q}^{T}_{ijk} \right]^{T}$, for all $i,j,k=1,2,3$.

\begin{lem}[\textbf{Lie group structure}]\label{lem3}
The tangent bundle $T\mathcal{S}^{3}$ is a Lie group under the binary operation \eqref{eq:GPct}, with
\begin{itemize}
    \item Group identity: $e = \left( \bar 1,0_{4\times 1}\right)\in T\mathcal{S}^{3}$, 
    \item Inverse: $g^{-1}=\left( q^{-1} , p^{-1}\right) \in T\mathcal{S}^{3}$, $\forall g=(q,p)\in T\mathcal{S}^{3}$, where $q^{-1}=\left[q_0,-\vec{q}^{T}\right]^{T}\in \mathcal S^3$, and $p^{-1}=\left[p_0,-\vec{p}^{T}\right]^{T}\in T_q S^3$.
\end{itemize}
\end{lem}
\begin{pf}
It is clear that $T\mathcal{S}^{3}$ is a smooth manifold and that the group operation \eqref{eq:GPct} is smooth. The group axioms are verified as follows:
\begin{enumerate}
\item $\forall g=(q,p)\in T\mathcal{S}^{3}$, it has
{\small \vspace{-3mm}
        \begin{align*}
            g\cdot e &= \Big( Q(q)\bar 1,\; Q(q)0_{4\times 1} + W(\bar 1)p +\lambda Q(q)( \bar 1 -\bar 1) \\
            &\quad +\lambda W(\bar 1)( q_{0}q-\bar 1) -\lambda ( q_{0}Q(q)\bar 1-\bar 1)  \Big) \\
            &= \Big( Q(\bar 1)q,\; Q( \bar 1)p + W( q)0_{4\times 1} +\lambda Q( \bar 1)( q_{0}\vec{q} -\bar 1) \\
            &\quad +\lambda W( q )( \bar 1-\bar 1) -\lambda ( q_{0}Q( \bar 1)q-\bar 1)  \Big) \\
            &= e\cdot g = \left( q,p\right) = g.
        \end{align*}}
    \item Let $g^{-1}=\left( q^{-1},p^{-1}\right)\in T\mathcal{S}^{3}$ be the inverse of $g=(q,p)\in T\mathcal{S}^{3}$, then
    {\footnotesize \vspace{-3mm}
        \begin{align*}
            & g\cdot g^{-1} = \\ 
            & \Big( Q(q)q^{-1} , \; Q(q)p^{-1} + W\left(q^{-1}\right)p +\lambda Q(q)(q_{0}q^{-1} -\bar 1)  \\
            &\quad\quad +\lambda W\left(q^{-1}\right) ( q_{0}q-\bar 1) -\lambda ( Q(q)q^{-1}-\bar 1) \Big) \\
            &= \Big( Q\left(q^{-1}\right)q , \; Q\left(q^{-1}\right)p + W(q)p^{-1} +\lambda Q\left(q^{-1}\right)(q_{0}q -\bar 1)  \\
            & \quad \quad +\lambda W(q)( q_{0}q^{-1}-\bar 1) -\lambda ( Q\left(q^{-1}\right)q-\bar 1)\Big)\\
            &= g^{-1}\cdot g =  \left(\bar 1, 0_{4\times 1}\right)= e,
        \end{align*}}
        where the fact that $q^{T}p=0$ is used, given that $p\in T_{q}\mathcal{S}^{3}$.
    \item Associativity:  $\forall g_{1}=(q_{1},p_1)$, $g_{2}=(q_{2},p_2)$ and $g_{3}=(q_{3},p_3)\in T\mathcal{S}^{3}$,
    {\small \vspace{-4mm}
\begin{align*}
    g_{1}&\cdot \left( g_{2}\cdot g_{3}\right) = g_{1}\cdot \Big( Q(q_{2})q_{3} , \; Q(q_{2})p_3 + W(q_{3})p_2 \\
    &+\lambda Q(q_{2})(q_{0,3}q_{3} -\bar 1) +\lambda W(q_{3})( q_{0,2}q_{2}-\bar 1) \\
    &-\lambda ( q_{0,23}Q(q_{2})q_{3}-\bar 1) \Big) \\
    =& \Big( Q(q_{1})Q(q_{2})q_{3},\; Q(q_{1})\Big( Q(q_{2})p_3 + W(q_{3})p_2 \\
    &+\lambda Q(q_{2})(q_{0,3}q_{3} -\bar 1) +\lambda W(q_{3})( q_{0,2}q_{2}-\bar 1) \\
    & -\lambda ( q_{0,23}Q(q_{2})q_{3}-\bar 1) \Big) +W\left( Q(q_{2})q_{3}\right) p_1 \\
    & +\lambda Q(q_{1})( q_{0,23}Q(q_{2})q_{3}-\bar 1) \\
    &+ \lambda W\left( Q(q_{2})q_{3}\right)( q_{0,1}q_{1}-\bar 1) \\
    & -\lambda ( q_{0,123}Q(q_{1})Q(q_{2})q_{3} - \bar 1) \Big)  \\
    =& \Big( Q\left( Q(q_{1})q_{2} \right)q_{3},\; Q\left( Q(q_{1})q_{2} \right) p_3 + W(q_{3})\Big( Q(q_{1})p_2 \\
    &+ W(q_{2})p_1 +\lambda Q(q_{1})(q_{0,2}q_{2} -\bar 1)  \\
    & +\lambda W(q_{2})( q_{0,1}q_{1}-\bar 1) -\lambda ( q_{0,12}Q(q_{1})q_{2}-\bar 1) \Big) \\
    &+\lambda Q\left( Q(q_{1})q_{2} \right) ( q_{0,3}q_{3} - \bar 1 ) \\ &+\lambda W(q_{3}) ( q_{0,12}Q(q_{1})q_{2}-\bar 1) \\
    &- \lambda ( q_{0,123}Q\left( Q(q_{1})q_{2} \right)q_{3} -\bar 1 ) \Big)\\
    =& \Big( Q(q_{1})q_{2} , \; Q(q_{1})p_2 + W(q_{2})p_1 \\
    &+\lambda Q(q_{1})(q_{0,2}q_{2} -\bar 1) +\lambda W(q_{2})( q_{0,1}q_{1}-\bar 1) \\
    &-\lambda ( q_{0,12}Q(q_{1})q_{2}-\bar 1) \Big) \cdot g_{3} \\
    =&\left(g_{1}\cdot g_{2}\right)\cdot g_{3},
\end{align*}}
where Properties of matrices $Q(\cdot)$ and $W(\cdot)$ in Lemma \ref{PtyQ} are used.
\end{enumerate}
\end{pf}

\subsection{Sliding Subgroup}

The sliding subgroup, defined below, inherits the structure of the Lie group $T\mathcal{S}^{3}$, allowing us to design sliding mode controllers on the Lie group.

\begin{lem}[\textbf{Sliding subgroup}]\label{lem4}
 For $g=(q,p)\in T\mathcal{S}^{3}$, let the sliding variable $s(g)$ be defined as
\begin{equation}\label{eq:s}
    s(g)= p+\lambda \left( q_0q-\bar 1\right) 
\end{equation}
where $\lambda>0$ is a scalar.

Then the set
\begin{equation}\label{eq:SubG}
    H = \left\{ g\in T\mathcal{S}^{3} \;| \; s(g) = 0_{4\times 1} \right\}
\end{equation}
is a Lie subgroup.
\end{lem}
\begin{pf}
Its clear that \eqref{eq:s} defines a smooth map. The associativity of $H$ is inherited from $T\mathcal{S}^{3}$. Therefore, it remains to verify the following: 
\begin{itemize}
    \item Identity $e\in H$:  $s(e)=s( \bar 1,0_{4\times 1}) = 0$. 
\item Inverse: Let $g=(q,p)\in H$. Then $s(g)=0\implies p=-\lambda \left( q_0 q-\bar 1\right)$. Then
\begin{align*}
    s \left( g^{-1} \right) &= p^{-1} + \lambda \left( q_0q^{-1} - \bar 1\right) \\
    &= - \lambda\left( q_0q^{-1} - \bar 1 \right) + \lambda \left( q_0q^{-1} - \bar 1\right) 
    = 0_{4\times 1}.
\end{align*}
 This shows that $g^{-1}\in H$.
\item Closure of the group operation: $\forall g_{1}=(q_{1},p_1)$ and $g_{2}=(q_{2},p_2)\in H$, it has that $p_1 = -\lambda\left( q_{0,1}q_{1}-\bar 1\right)$ and  $p_2 = -\lambda\left( q_{0,2}q_{2}-\bar 1\right)$. Thus, 
{\small \vspace{-3mm}
\begin{align*}
    s\left( g_{1}\cdot g_{2}\right) &= s\Big( Q(q_{1})q_{2} , \; Q(q_{1})p_2 + W(q_{2})p_1 \\
    &\quad +\lambda Q(q_{1})(q_{0,2}q_{2} -\bar 1) +\lambda W(q_{2})( q_{0,1}q_{1}-\bar 1) \\
    &\quad -\lambda ( q_{0,12}Q(q_{1})q_{2}-\bar 1) \Big) \\
    &= Q(q_{1})p_2 + W(q_{2})p_1 +\lambda Q(q_{1})(q_{0,2}q_{2} -\bar 1) \\
    &\quad +\lambda W(q_{2})( q_{0,1}q_{1}-\bar 1) -\lambda ( q_{0,12}Q(q_{1})q_{2}-\bar 1) \\
    &\quad +\lambda ( q_{0,12}Q(q_{1})q_{2} - \bar 1) \\
    &= Q(q_{1})( -\lambda( q_{0,2}q_{2}-\bar 1) ) \\
    &\quad + W(q_{2})( -\lambda( q_{0,1}q_{1}-\bar 1) ) \\ &\quad +\lambda Q(q_{1})(q_{0,2}q_{2} -\bar 1) +\lambda W(q_{2})( q_{0,1}q_{1}-\bar 1) \\
    &= 0_{4\times 1}. 
\end{align*}}
Therefore,  $g_1\cdot g_2\in H$.
\end{itemize}
\end{pf}

\subsection{Sliding Property of the Sliding Subgroup}
One of the salient features of sliding mode control is that the behavior of a system on the sliding subgroup is independent of the system dynamics. Define $\mathcal{M}\vcentcolon = \left\{ g=(q,p)\in T\mathcal{S}^{3} \; |\; p=\dot{q}  \right\}$, then $\mathcal{M}\subset T\mathcal{S}^{3}$. The following lemma establishes the sliding property of the subgroup $\bar{H}\vcentcolon =  H \cap \mathcal{M}$, where $H$ is defined in \eqref{eq:SubG}, that is, $\bar{g}\to e$ asymptotically provided that $\bar{g}\in \bar{H}$.

\begin{lem}[\textbf{Sliding property}]\label{lem5}
Let $\bar{g}=(q,\dot{q})\in \bar{H}$, then $\bar{g}\to e=(\bar{1},0_{4\times 1})$ asymptotically. 
\end{lem}
\begin{pf}
Since $\left( q,\dot{q}\right)\in H \; \implies \; s(\bar{g})=0_{4\times 1}$, it follows that 
\begin{equation*}
  \frac{d}{dt} ( q - \bar 1) =  \dot{q} =  -\lambda ( q_0q - \bar 1).
\end{equation*}

Consider the candidate Lyapunov function $V=\frac{1}{2}\| q-\bar 1 \|^{2}=1-q_0$. Its time derivative on the sliding subgroup is 
\begin{align*}
    \dot{V} &= ( q - \bar 1)^{T}\dot{q} = ( q - \bar 1)^{T}\left( -\lambda( q_0q - \bar 1) \right)  \\
    &= \lambda \bar 1^{T}( q_0q - \bar 1) = -\lambda( 1 - q_0^{2})=-\lambda (1+q_0)V.
\end{align*}
For all $q\in\mathcal{S}^{3}\backslash \{-\bar 1\}$, it has $1+q_0< 0$. Therefore,  $\dot V<0$. This shows the almost global asymptotic stability of $q=\bar 1$. 
\end{pf}

\subsection{Stability of the Sliding Subgroup}
Consider the Lagrangian dynamics \eqref{eq:EL-model} with the configuration manifold $\mathcal{M}$. Let $\bar{g}=\left( q,\dot{q}\right)\in \mathcal{M}$, then the following control law 
\begin{equation}\label{eq:bTauCtrl}
    \bar{\tau} = -\lambda\left( D(q)(q_0\dot{q} + \dot{q_0}q) +C(q,\dot{q})( q_0q - \bar 1) \right)-K_{r}s( \bar{g})
\end{equation}
renders the sliding subgroup $\bar{H}$ attractive, where $K_{r}=K^{T}_{r}\in\mathbb{R}^{4\times 4}$ is a positive definite gain matrix.

\begin{lem}[\textbf{Stability of the Sliding Subgroup}]\label{lem6}
The control law \eqref{eq:bTauCtrl} in closed loop with the system \eqref{eq:EL-model} renders the equilibrium point $s( \bar{g})= 0_{4\times 1}$ globally exponentially stable.
\end{lem}
\begin{pf}
By the definition of the sliding variable \eqref{eq:s} for $s( \bar{g})$ and the Lagrangian dynamics \eqref{eq:EL-model}, it has 
\begin{align*}
    D(q)\dot{s} &= D(q)\ddot{q} + \lambda D(q) (q_0\dot{q} + \dot{q_0}q),\\
    &= -C(q,\dot{q})\dot{q} +\bar{\tau} + \lambda D(q)(q_0\dot{q} + \dot{q_0}q).
\end{align*}
In closed loop with the control law \eqref{eq:bTauCtrl} it yields
\begin{equation}\label{eq:srp}
    D(q)\dot{s} = -C(q,\dot{q}) s - K_{r}s.
\end{equation}

Consider the following Lyapunov function candidate 
\begin{equation}\label{eq:Vr}
    V_{r}(s) = \frac{1}{2}s^{T}D(q)s,
\end{equation}
which is positive definite and radially unbounded
according to Property \ref{pLg1} of Lemma \ref{lem2}.

The time derivative of \eqref{eq:Vr} along the trajectories of system \eqref{eq:srp} is 
\begin{align}\label{eq:Vrp}
    \dot{V}_{r} &= s^{T}D(q)\dot{s} + \frac{1}{2}s^{T}\dot{D}(q)s \notag\\
    &= s^{T}\left(  -C(q,\dot{q}) s - K_{r}s\right) + \frac{1}{2}s^{T}\dot{D}(q)s \notag\\
    &= -s^{T}K_{r}s \notag\\
    &\leq -\lambda_{\min}\left(K_{r}\right)\|s\|^{2} \notag\\
    &\leq -2\frac{\lambda_{\min}\left(K_{r}\right)}{\lambda_{\max}(M)} V_{r},
\end{align}
where the skew-symmetry property \eqref{pLg2} of Lemma \ref{lem2} is used. Therefore, 
\begin{equation*}
    \|s(t)\|\leq \left(\frac{\lambda_{\max}(M)}{\lambda_{\min}(M)}\right)^{1/2}\|s(t_{0})\|\mathrm{e}^{-\frac{\lambda_{\min}(K_r)}{\lambda_{\max}(M)}(t-t_{0})},
\end{equation*}
giving that $s\to 0_{4\times 1}$ exponentially (Theorem 4.10, \cite{hassan2002nonlinear}). 
\end{pf}

\begin{rem}\label{rmk1}
 Due to the double coverage of $SO(3)$ by $\mathcal S^3$,  both $\pm q$ will give the same attitude. Thus, the sliding variable \eqref{eq:s} can be changed to $s(\bar{g})= \dot{q}+\lambda \left( q_0q\pm\bar 1\right)$ according to the sign of the scalar part $q_0$ of the unit quaternion. Furthermore, to ensure robustness to arbitrarily small noise in measurements, a switching strategy may be used in the control law \citep{mayhew2011quaternion}. For the same reason, the global exponential stability established on the group $T\mathcal{S}^{3}$ must be interpreted with caution, since there are no continuous control laws that achieve global stability on any compact group due to topological constraints \citep{bhat2000topological}.
\end{rem}

\section{Attitude tracking control}  \label{sec:sec4}

\subsection{Control Objectives}
Given a desired trajectory $\dot{q}_{d}=\frac{1}{2}J(q_{d})\omega_{d}$, where $\omega_{d}\in\mathbb{R}^{3}$ is a desired angular velocity that is twice differentiable and $q_{d}\in\mathcal{S}^{3}$ is the desired attitude, the goal of tracking the attitude is to render $\left( q,\dot{q} \right) \to \left( q_{d},\dot{q}_{d} \right)$ asymptotically. 

Let $g_{d}:=\left(q_{d},\dot{q}_{d}\right)$, $g:=(q,\dot{q})\in \mathcal M$. Then the intrinsic error defined on the Lie group $T\mathcal{S}^{3}$ is defined as
\begin{align}\label{eq:gErr}
    g_{e}&=g^{-1}_{d}\cdot g  = \left( q_{e}, p_{e} \right)\in T\mathcal{S}^{3} ,\\
    q_e &= Q( q^{-1}_{d})q = Q^T(q_d)q , \label{eq:qErr} \\
    p_e &= Q( q^{-1}_{d}) \dot{q} + W(q)\dot{q}^{-1}_{d} +\lambda Q( q^{-1}_{d})(q_0q -\bar 1) \notag \\
    &\quad +\lambda W(q)( q_{0,d}q^{-1}_{d}-\bar 1) -\lambda \left( ( q^{T}_{d}q)Q( q^{-1}_{d})q-\bar 1\right) .
\end{align}
Note that by Eq. \eqref{eq:KinQ}, 
\begin{equation}\label{eq:qErrp}
    \dot{q}_{e} = Q( q^{-1}_{d}) \dot{q} + W(q)\dot{q}^{-1}_{d} = \frac{1}{2}J(q_{e})\omega_{e},
\end{equation}
where $ \omega_{e} = \omega - R^{T}(q_{e})\omega_{d}$
is the angular velocity error.

Therefore, $\bar{g}_{e} = (q_{e},\dot{q}_{e})\in \mathcal{M}$ describes the intrinsic error on the tangent bundle of the Lagrangian system
\begin{equation}\label{eq:ER-model}
    D(q_{e})\ddot{q}_{e} + C(q_{e},\dot{q}_{e})\dot{q}_{e} = \bar{\tau}_{c},
\end{equation}
with
\begin{align}
    D(q_{e}) &= J(q_{e})MJ^{T}(q_{e}) + m_{0}q_{e} q_{e}^{T}, \label{eq:Dveps}\\
    C(q_{e},\dot{q}_{e}) &= -J(q_{e})w^{\wedge}_rJ^{T}(q_{e})-D(q_{e})Q(\dot{q}_{e})Q^{T}(q_{e}), \label{eq:Cveps}\\
    \bar{\tau}_{c} &= \frac{1}{2}J(q_{e})\tau_{c} ,\label{eq:bTff}
\end{align}
where 
\begin{align}
    w_r &= M\omega_{e} - \left(\mathrm{tr}(M)I_{3} - 2M  \right) R^{T}(q_{e})\omega_{d} \label{eq:w},\\
    \tau_{c} &= S\left( MR^{T}(q_{e})\omega_{d} \right) R^{T}(q_{e})\omega_{d} - MR^{T}(q_{e})\dot{\omega}_{d} + \tau . \label{eq:Tff}
\end{align}

Therefore, it follows from the results of the previous section that the control objectives are achieved by designing a control law for $\bar{\tau}_{c}$ that makes the sliding subgroup $\bar{H}$ attractive. Subsequently, $\bar{g}_{e}\to e$ as a consequence of the sliding property.

\subsection{Controller Design}

The intrinsic error $g_{e}\in T\mathcal{S}^{3}$ can be expressed in the state manifold of the Lagrangian system \eqref{eq:ER-model} as $\bar{g}_{e}=(q_{e},\dot{q}_{e})\in \mathcal M$. The sliding variable therefore, according to \eqref{eq:s}, is given by
\begin{equation}\label{eq:s1}
    s\left( \bar{g}_{e}\right) = \dot{q}_{e} + \lambda ( q_{0,e}q_{e} - \bar 1),
\end{equation}
$\lambda>0$ being a design parameter. The following control law will render the sliding subgroup $\bar{H}$ attractive
\begin{align}
    \bar{\tau}_{c} &= -\lambda\left( D(q_{e})\left(q_{0,e}\dot{q}_{e} + \dot{q}_{0,e}q_{e}\right) +C(q_{e},\dot{q}_{e})\left( q_{0,e}q_{e} - \bar 1\right) \right) \notag \\
    &\quad -K_{r}s\left( \bar{g}_{e}\right), \label{eq:CtrlS1}
\end{align}
where 
$K_{r}=K^{T}_{r}\in\mathbb{R}^{4\times 4}$ is a positive definite gain matrix. 

\begin{thm}[\textbf{Almost-global asymptotic stability}]\label{thm1}
The control law \eqref{eq:CtrlS1} in closed loop with the Lagrangian dynamics \eqref{eq:ER-model} stabilizes the equilibrium point $ \bar{g}_{e} = e$ almost globally asymptotically.
\end{thm}
\begin{pf}
It follows directly of the sliding variable $s(\bar{g}_{e})$ as a consequence of Lemma \ref{lem4}, Lemma \ref{lem5} and Lemma \ref{lem6}.
\end{pf}



\section{SIMULATIONS}  \label{sec:sec5}

To illustrate the theoretical results and the performance of the GSMC, the controller \eqref{eq:CtrlS1} was compared with the controller (37) in \cite{espindola2023four} developed in Euclidean space $\mathbb R^4$. It was shown in \cite{espindola2023four} that the sliding surface
\begin{equation}\label{eq:s0}
    s_{0} = \dot{q}-\dot{q}_{d} + \Lambda\left( q-q_{d}\right) \in \mathbb R^4
\end{equation}
is exponentially stable, $\Lambda=\Lambda^T>0$ being a gain matrix.

The desired trajectory was given by $\omega_{d}(t) = [0,0,0.1]^{T}$ ($\mathrm{rad}/s$). Initial conditions and design parameters are shown in Table \ref{tab:Parameters}. The inertia matrix considered is given by 
\begin{equation}\label{eq:M}
    M = \left[ \begin{array}{ccc}
         3.6046 & -0.0706 & 0.1491\\
         -0.0706 & 8.6868  & 0.0449 \\
         0.1491  & 0.0449 & 9.3484
    \end{array}\right] \; \left(\mathrm{Kg}\mathrm{m}^{2}\right).
\end{equation}

For comparison purposes, the same controller gains were chosen for both controllers under the ideal scenario, where the measurements were noise-free and the inertia matrix \eqref{eq:M} was known. For further comparisons under non-ideal situations, simulations under uncertainties of $\pm 30 \%$ in the inertia matrix $M$, and in the presence of measurement noises were carried out, respectively. 

\begin{center}
\begin{table}[t]%
\centering
\caption{Parameters and initial conditions.\label{tab:Parameters}}%
\begin{tabular*}{240pt}{@{\extracolsep\fill}lll@{\extracolsep\fill}}
\toprule
\textbf{Initial Condition} & \textbf{Value}  & \textbf{Units} \\
\midrule
$q(0)$ & $\left[0,\bar{u}^{T}\right]^{T}$\tnote{$\dagger$} & \\
$\omega(0)$ & $\left[0,0,0\right]^{T}$ & $\mathrm{rad}/s$ \\
 $q_{d}(0)$ & $\left[1,0,0,0\right]^{T}$ & \\
$\omega_{d}(0)$ & $\left[0,0,0.1\right]^{T}$ &  $\mathrm{rad}/s$ \\ \midrule
\textbf{Parameter} & \textbf{Value}  & \textbf{Units} \\ \midrule
$m_{0}$ & $6$ & $\mathrm{Kg}\mathrm{m}^{2}$ \\ 
 $K_{r}$ & $3I_{4}$ & \\
 $\lambda$ & $0.1$ & \\ \midrule
\textbf{ \cite{espindola2023four} }& \textbf{Value}  & \textbf{Units}\\ \midrule
$K_{s}$ & $3I_{4}$ & \\
 $\Lambda$ & $0.1 I_{4}$ &\\
\bottomrule
\end{tabular*}
\begin{tablenotes}
\item[$\dagger$] $u\vcentcolon=[1,2,3]^{T}$, and $\bar{u}=u/\|u\|$.
\end{tablenotes}
\end{table}
\end{center}

\begin{figure}[t]
       \centering
       \includegraphics[trim = 10mm 5mm 20mm 0mm,scale=0.28]{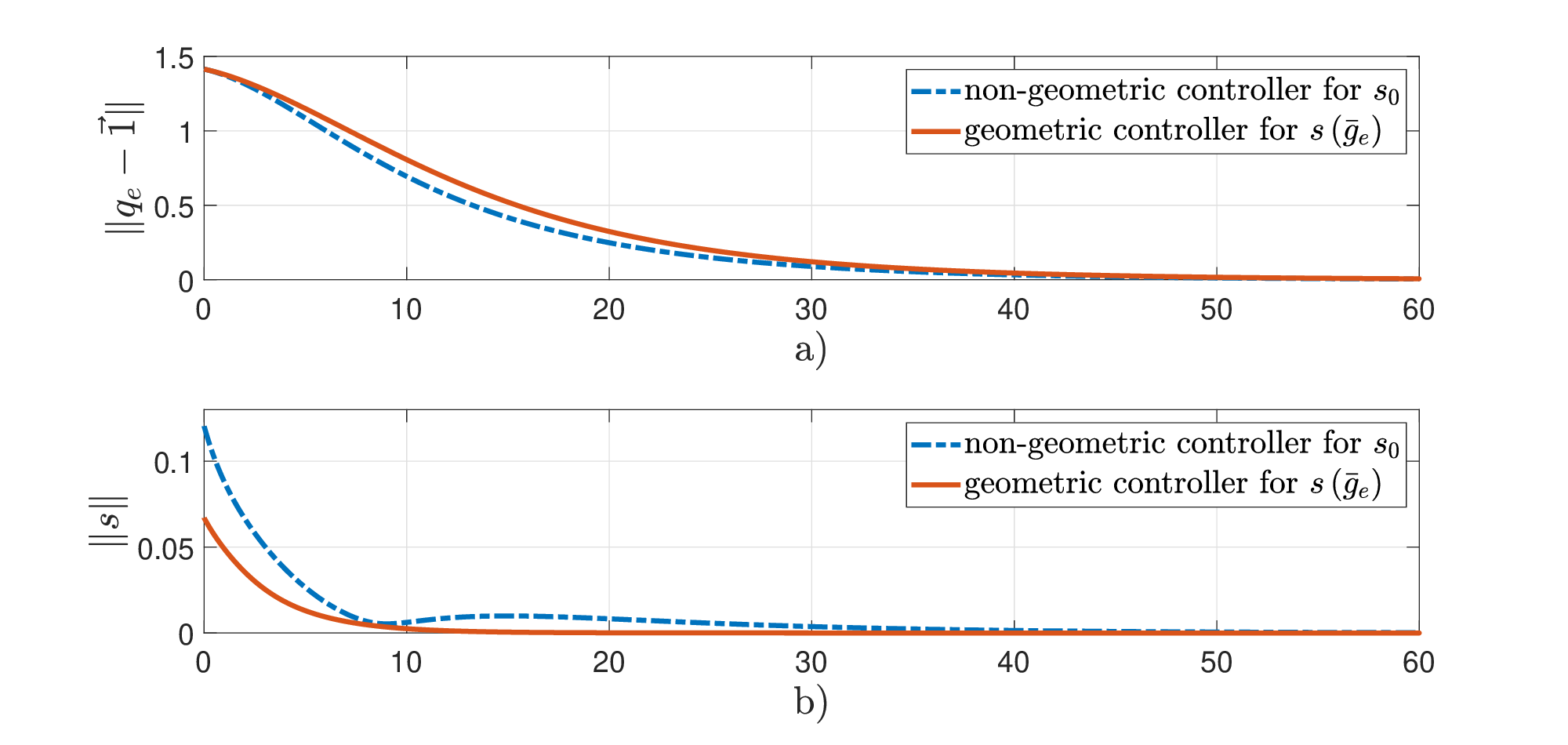}
		\includegraphics[trim = 10mm 5mm 20mm 0mm,scale=0.28]{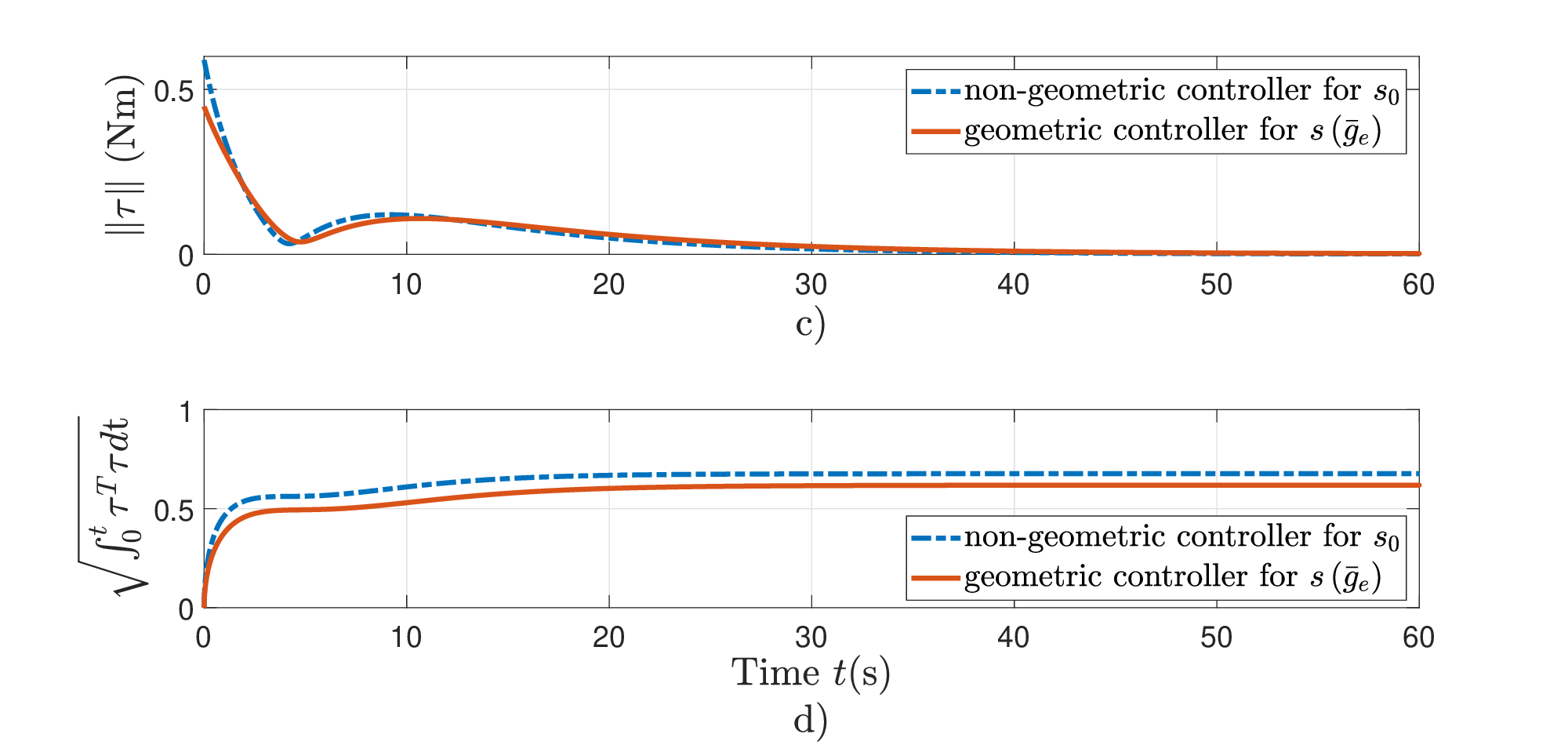}
       \caption{The ideal case:  Performance comparison of the non-geometric controller in \cite{espindola2023four} 
       (blue line) and the geometric controller \eqref{eq:CtrlS1} (red line).}
       \label{fig:IdCase}
\end{figure}

Fig.  \ref{fig:IdCase} illustrates the behavior of the controllers in the ideal situation. Fig. \ref{fig:IdCase}(a) shows the norm $\| q_{e}-\bar 1 \|$. Note that both controllers achieved the convergence at $50$ $(s)$. Fig. \ref{fig:IdCase}(b) shows the norm of the sliding variables $\|s\left( \bar{g}\right)\|$ and $\|s_0\|$. A large transient response in $s_{0}$ of the non-geometric controller is observed,  which caused it to take a longer time ($40$ (s)) to converge, while the geometric controller stabilizes $s\left( \bar{g}\right)$ at $15$ ($s$). 
Fig.  \ref{fig:IdCase}(c) draws the norm of the torque $\|\tau\|$. Note that the non-geometric controller requires slightly higher torque input than the geometric controller at the transient. 
Finally, Fig. \ref{fig:IdCase}(d) illustrates the energy consumption of each controller in terms of $\sqrt{\int^{t}_{0}\tau^{T}\tau d\mathrm{t}}$. The lower energy consumption in the geometric controller can be noticed. 

\begin{figure}[t]
       \centerline{
       \includegraphics[trim = 10mm 0mm 20mm 0mm,scale=0.28]{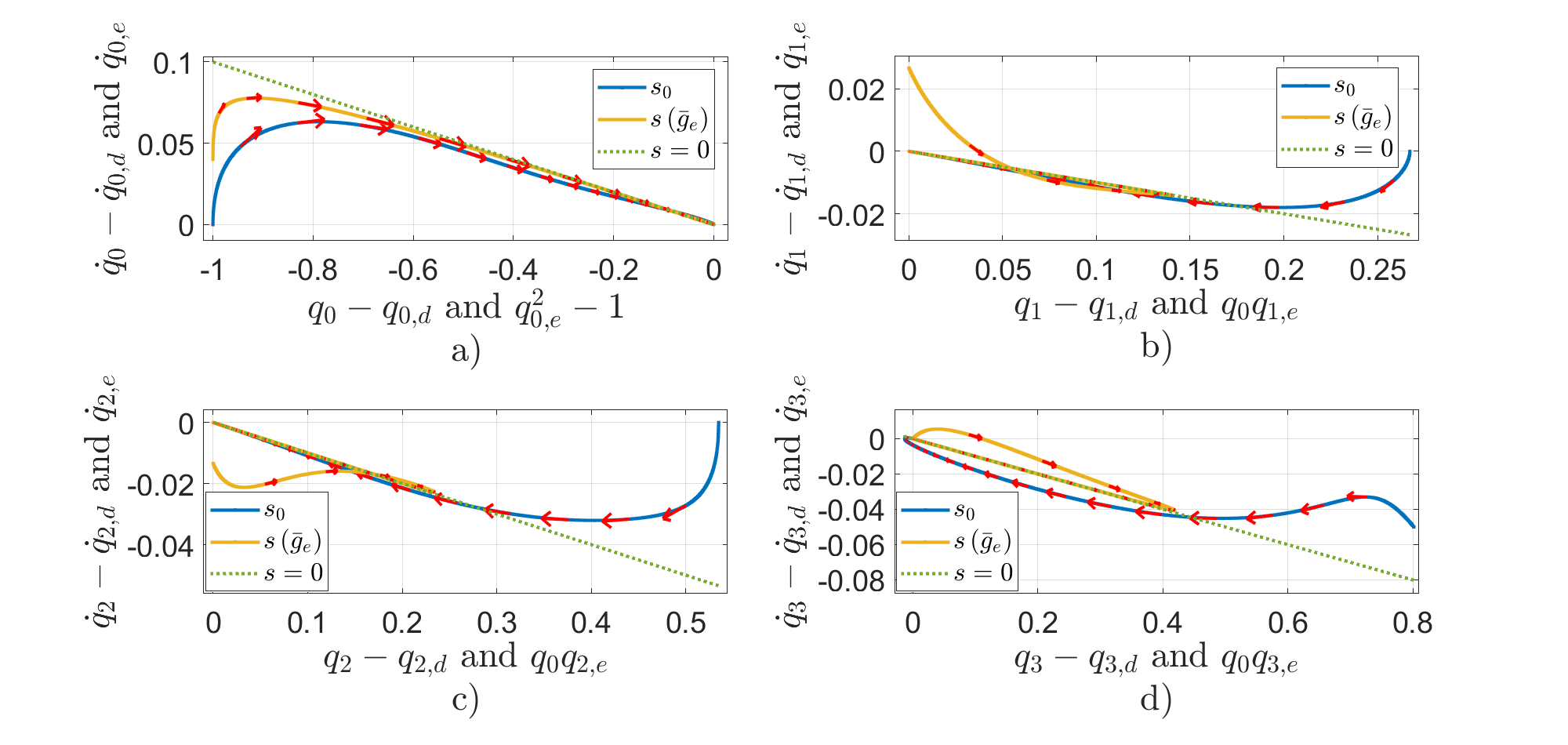}}
       \caption{The ideal  case:  behaviour of the sliding variables in both controllers.}
       \label{fig:SldS}
   \end{figure}

   To better appreciate the convergence behavior of these controllers, the sliding variable of both controllers is shown in Fig. \ref{fig:SldS}. Note that the behavior of the sliding surface $s\left( \bar{g}_{e}\right)$ of the geometric controller first exhibits the convergence to the siding surface $s=0_{4\times 1}$ and later to the equilibrium $\left( s\left( \bar{g}_{e}\right),q_{e}\right) = \left( 0_{4\times 1}, \bar 1 \right)$. On the other hand, the sliding surface $s_{0}$ of the non-geometric controller  oscillates around the surface $s_0=0_{4\times 1}$ and then convergences to the equilibrium $\left( s_0,\ q_{e}\right) = \left( 0_{4\times 1}, \bar 1 \right)$ with an oscillating behavior, leading to more energy consumption than the geometric controller.

\begin{figure}[t]
       \centering
       \includegraphics[trim = 10mm 5mm 20mm 0mm,scale=0.28]{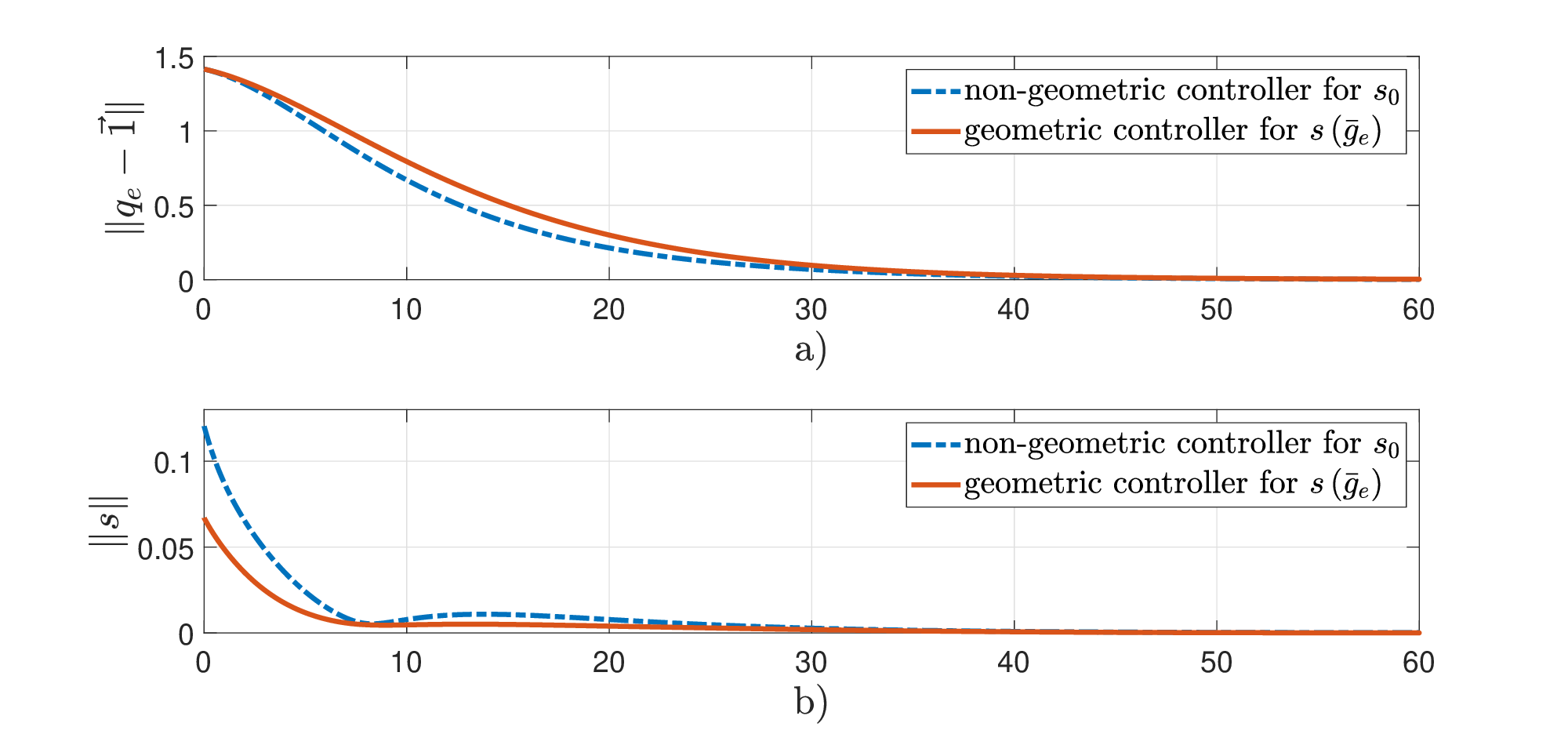}
		\includegraphics[trim = 10mm 5mm 20mm 0mm,scale=0.28]{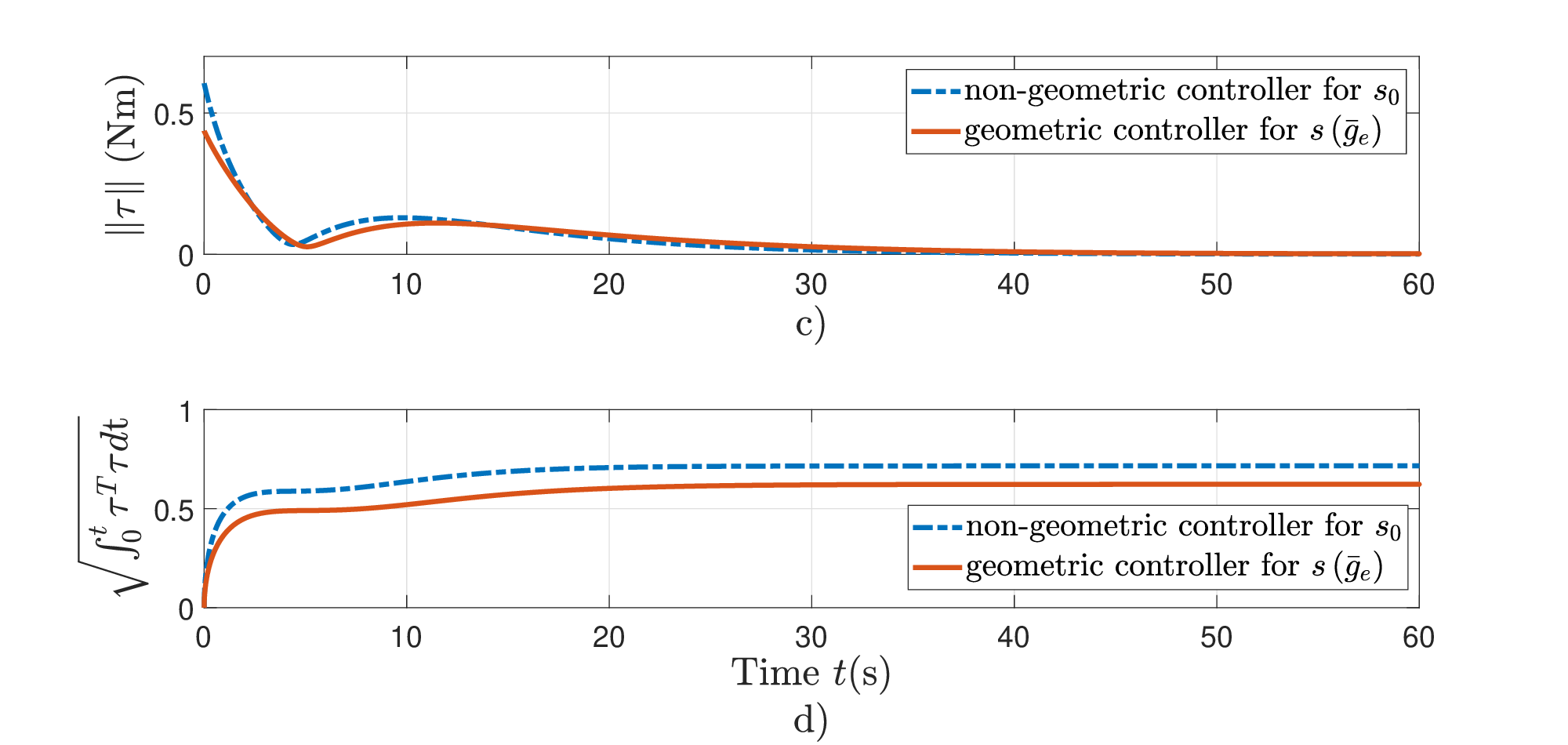}
       \caption{Uncertain inertia matrix:  Performance of the  non-geometric controller in \cite{espindola2023four}  (blue line) and the geometric controller \eqref{eq:CtrlS1} (red line).}
       \label{fig:UpCase}
   \end{figure}

In the presence of uncertainties of $\pm 30 \%$ in the inertia matrix, Fig. \ref{fig:UpCase} shows the performance of both controllers. Notice from Fig. \ref{fig:UpCase}(a)-(d) that the performance of both controllers is similar to that in the ideal case, being the geometric controller \eqref{eq:CtrlS1} more energy efficient than the non-geometric controller (Fig. \ref{fig:UpCase}(d)). 
   
\begin{figure}[t]
       \centering
       \includegraphics[trim = 10mm 5mm 20mm 0mm,scale=0.28]{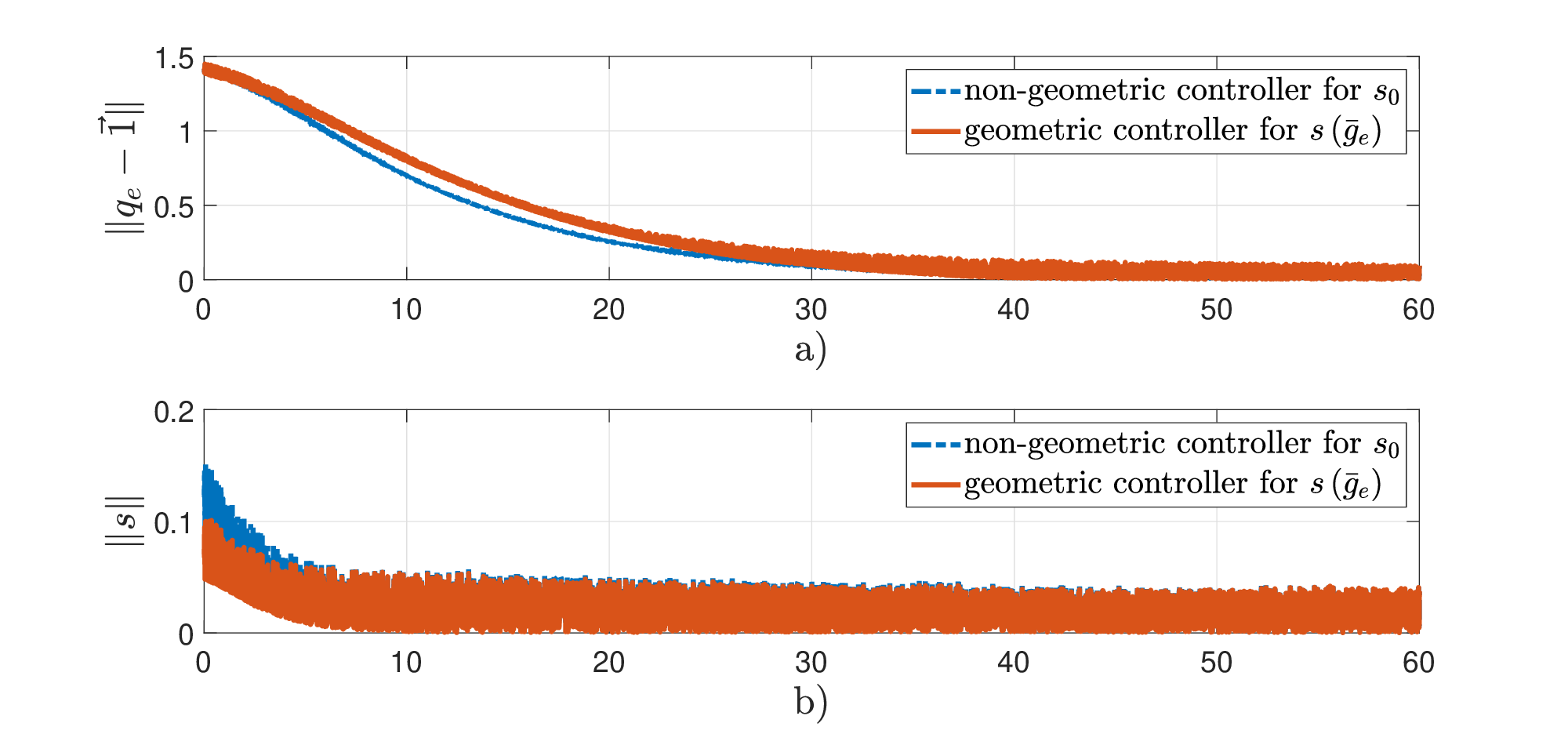}
		\includegraphics[trim = 10mm 5mm 20mm 0mm,scale=0.28]{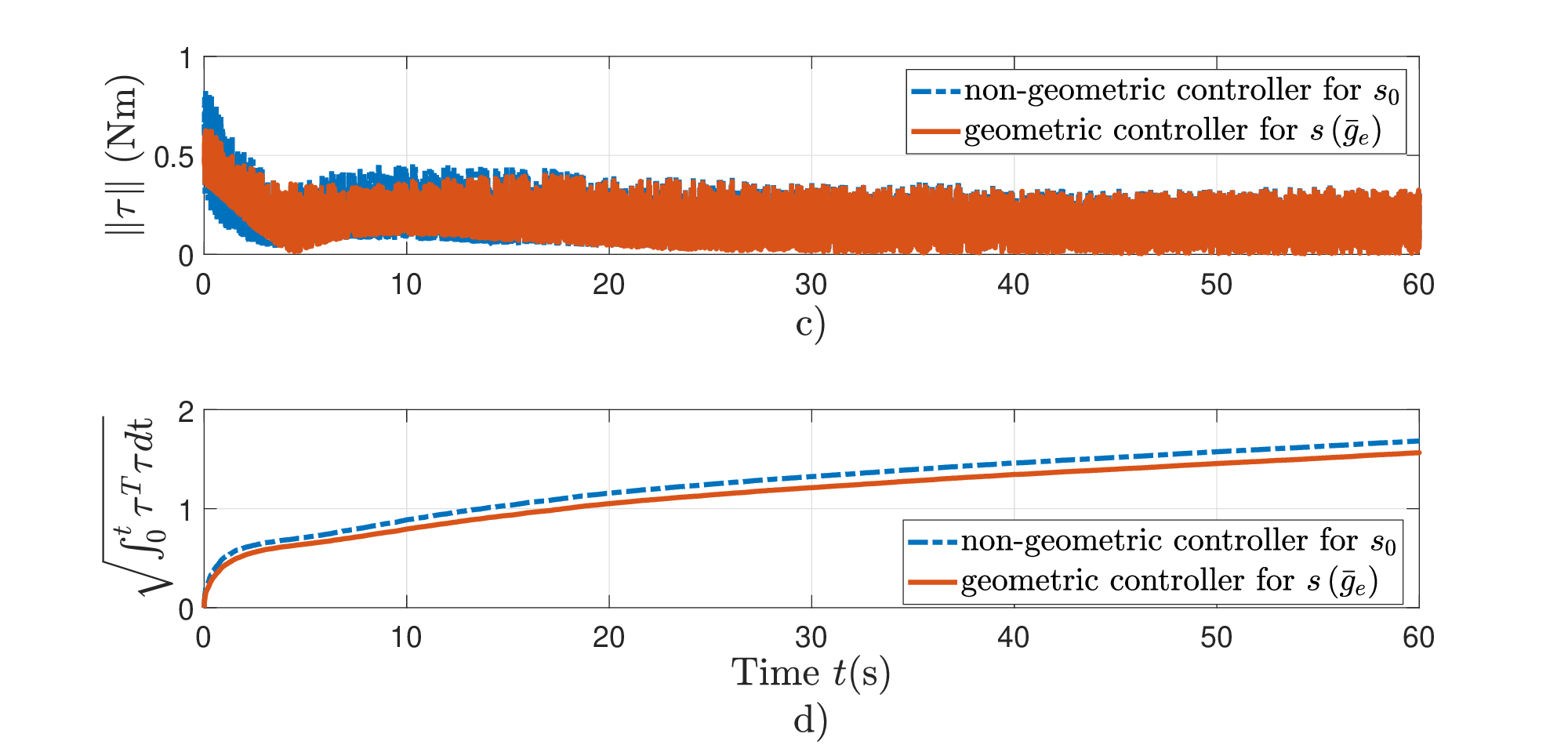}
       \caption{Noisy measurements: Performance of the  non-geometric controller in \cite{espindola2023four}  (blue line) and the geometric controller \eqref{eq:CtrlS1} (red line).}
       \label{fig:NysCase}
   \end{figure}

In the last simulation, noise in the quaternion and the angular velocity  measurements were added by $q_m=(q+n_1\nu)/\|q+n_1\nu\|$ and $\omega_m=\omega+n_2 u$, where $\nu\in\mathbb R^4$ and $u\in\mathbb R^3$ are vectors with  Gaussian distribution entries of zero mean and unit variance, $n_1\in (0, 0.1)$ and $n_2\in (0, 0.1)$  are uniform distributions. Fig. \ref{fig:NysCase} illustrates the behavior of both controllers under these noisy measurements. Note that the norm $\|q_e-\bar 1\|$ oscillates inside $[0, 0.1]$  in the steady state (Fig. \ref{fig:NysCase}(a)), which corresponds to the noise variance of $q_m$. Likewise, the norm of the sliding variable was kept inside $[0, 0.05]$ after the transient in both controllers (Fig. \ref{fig:NysCase}(b)), which is smaller than the noise variance of the angular velocity measurement $\omega_m$. 
However, a larger impact of the noise on the behavior was observed in the non-geometric controller of \cite{espindola2023four}, which caused  a larger torque input and more energy consumption than in the geometric  (Figs. \ref{fig:NysCase}(c)-(d)).

\section{Conclusions}\label{sec:sec6}
A geometric sliding mode controller on the quaternion group was developed in this paper for attitude tracking. The state manifold of the Lagrangian dynamics that describes the attitude using the unit quaternion was shown to be a Lie group. A sliding subgroup that heritages the Lie group structure was designed, which allows us to design a sliding mode control on the Lie group. The almost global asymptotic stability of the closed-loop system was achieved. The performance of the proposed geometric sliding mode controller was illustrated and compared with a non-geometric sliding mode controller in the simulations, showing that the geometric controller performed better than the non-geometric controller in the presence of uncertain inertia matrix and noise in the measurement. Furthermore, more energy efficiency was observed in the geometric controller.  


\section*{Acknowledgments}
This work was partially supported by PAPIIT-UNAM IN112421. 







\appendix
\section{Properties of the Maps}\label{app1}

Some properties of map $J$ are listed below.

\vspace{10pt}
\emph{\textbf{Properties of map $J$}} \citep{espindola2023four}:\label{PropJx}
For all $x, y\in \mathbb{R}^{4}$, the following properties hold for  $J(\cdot)$:
\begin{enumerate}
\item $J^{T}(x)y = -J^{T}(y)x$, \label{pJ1}
\item $J^{T}(x)y = 0_{3\times 1}$, for all $y = kx,\; \forall k\in\mathbb{R}$, \label{pJ2}
\item $J^{T}(x)J(x)=||x||^{2} I_{3}$. \label{pJ3}
\end{enumerate}
These same properties also hold for map $L$.

Likewise, some properties of maps $Q$ and $W$ are summarized in the following lemma.

\begin{lem}[\textbf{Properties of maps} $Q(x)$ and $W(x)$]\label{PtyQ}
Maps $Q(x),W(x): \mathbb R^4\to \mathbb{R}^{4\times 4}$ defined in  \eqref{eq:Q} verify the following properties $\forall x,y\in\mathbb{R}^{4}$
\begin{enumerate}
\item $Q(x)\in SO(4), \; \forall x\in \mathcal{S}^{3}$, \label{pQ1}
\item $Q(y)Q^T(x) = J(y)J^T(x) + yx^{T}$, \label{pQ2}
\item $u^{T}Q(y)Q^T(x)u = 0 \iff y^{T}x = 0$, $\forall u\in\mathbb{R}^{4}$, \label{pQ3}
\item $Q(Q(x)y) = Q(x)Q(y)$, \label{pQ4}
\item $Q\left(Q^T(x)y\right) = Q^T(x)Q(y)$, \label{pQ5}
\item $Q(\bar{w})\in\mathfrak{so}(4)$, $\forall \bar{w}=[0,w^{T}]^{T}\in\mathbb{R}^{4}$, $w\in\mathbb{R}^{3}$, \label{pQ6}
\item $Q(x)y=W(y)x$, \label{pQ7}
\item $Q(x)W(y) = W(y)Q(x)$, \label{pQ8}
\item $W\left( Q(x)y\right) = W(y)W(x)$.\label{pQ9}
\end{enumerate}
\end{lem}
\begin{pf}
Properties \ref{pQ1}-\ref{pQ3} are proved in \cite{espindola2023four}. Properties \ref{pQ4}-\ref{pQ5} and \ref{pQ7}-\ref{pQ9} can be verified by direct substitution. Finally, Property \ref{pQ6} is trivial by  considering $x_0=0$ in \eqref{eq:Q}.
\end{pf}

\clearpage
\bibliographystyle{agsm}            

\bibliography{geo1}%




\end{document}